\documentclass[twocolumn,showpacs,preprintnumbers,amsmath,amssymb]{revtex4}
\usepackage{epsfig} % use alternative interface for graphics similar to epsf
\usepackage{dcolumn}% Align table columns on decimal point
\usepackage{bm}% bold math

\begin{document}

\draft

%\wideabs{

\title{Analysis of Hadronic Properties at SPS energies from a Statistical Model}
\author{Kang Seog Lee }
\address{   Department of Physics, Chonnam National University,
Gwangju 500-757, Korea}

\date{Jan. 8, 2004}

\begin{abstract}

Using a statistical model of multiparticle production by Chou,
Yang and Yen, which looks similar to the thermal model and is
known to account for the single particle distributions in e$^+$
e$^-$ collisions, we fit the rapidity and transverse mass spectra
of pions and kaons measured in Pb+Pb collisions by NA49
collaboration. This model nicely fits both the rapidity and the
transverse mass spectra of each particle species  with a few
parameters. However, fitting all the particles with a single value
of $T_p$ is not possible. This analysis shows that the success of
the thermal models does not necessarily mean the thermalization of
the system in collision.

\end{abstract}


\maketitle

%\narrowtext
%

%%%%%%%%%%%%%%%%%%%%%%%%%%%%%%%%%%%%%%%%%%%%%%%%%%%%%%%%
\section{INTRODUCTION}
\label{sec:intro}
%%%%%%%%%%%%%%%%%%%%%%%%%%%%%%%%%%%%%%%%%%%%%%%%%%%%%%%%

It has been of great interest to find out whether a hot and dense
nuclear matter has been formed during the high energy nuclear
collisions\cite{qm}, and when the temperature and/or density is
high enough one expects a system of free quarks and gluons, namely
quark-gluon plasma. Main issue here is the thermalization of the
partonic or hadronic system formed just after the collisions and
throughout the evolution of the following hadronic system. Thus
probing the thermalization of the system in collision is of great
importance. The large elliptic flow and the jet quenching effect
indicate a collective behavior. However those are not yet the
direct experimental evidence of thermalization. Best one can do so
far is to assume thermalization and to show the experimental data
is not inconsistent with the assumption of thermalization. In this
case one has to show that assuming thermalization is the only way
to explain the data in order to claim the thermalization. We will
show in this paper that a simple statistical model can fit both
the rapidity and the transverse momentum spectra of various
hadrons.

In the relativistic heavy-ion collisions at AGS, SPS and RHIC,
hadrons are measured quite accurately, especially for pions, kaons
and protons. Usually the hadronic data are presented in many
different ways: (1) total multiplicities of particles or the
ratios of the multiplicities of different hadrons, (2) single
differentials such as the rapidity distribution $dN/dy$, and the
transverse mass spectra $dN / m_{\perp} d m_{\perp} $ , (3) the
double differentials such as B-E correlation function and (4) the
averages or moments of certain physical quantities such as
fluctuations and flow coefficients which is the second moment of
the azimuthal angle, $\phi$. The total number of particles can be
obtained either by integrating the rapidity distribution over the
whole range of the rapidity, {\it i.e.} $N_i = \int dy (dN_i /dy
)$ or by integrating over the whole $p_{\perp}$ range, {\it i.e.}
$N_i = \int m_{\perp} d m_{\perp} dN / m_{\perp} d m_{\perp}$.

Models assuming thermal and chemical equilibrium fit each pieces
of data very nicely\cite{braun, becattini,rafelski,lee,
heinz,broniowski} and explain many features of data. Thermal
models assuming chemical equilibrium\cite{braun,
becattini,rafelski}, when applied to the ratios of various hadron
multiplicities fit the data impressively well with only a few
parameters such as the temperature $T_c$, baryonic chemical
potential, $\mu_b$ and/or the strange chemical potential, $\mu_s$.

Fireball models with strong radial flow\cite{lee,heinz,broniowski}
fits the transverse mass spectra quite nicely with a few
parameters, $T_{th}$, $\mu_b$ and radial expansion velocity at the
surface, $\beta_s$.

The rapidity distributions are usually Gaussian. Spherically
expanding fireball model applied to pion data at AGS
energy\cite{lee} results in too narrow width compared to the data,
and in the cylindrically expanding fireball model\cite{heinz} fast
longitudinal expansion makes the rapidity width wider and seems to
fit the data when the system sizes are small. However,  expanding
fireball models with any geometry seems to have difficulty in
fitting  the rapidity distributions of pions and kaons,
simultaneously, in the Pb+Pb collision at SPS.

The two different sets of freeze-out parameters from the ratio
analysis and  the transverse momentum spectra analysis leads to
the conjecture that the chemical components freezes earlier at
high temperature, $T_{ch}$, and the thermal equilibrium is still
maintained until $T_{th}$, keeping the number of certain particle
species constant\cite{redlich}.

However, calculation by Broniowski and
Florkowski\cite{broniowski} claim that both the ratios and
transverse mass spectra can be fitted with a single set of
freeze-out parameters. As the models seem to be the similar,
independent calculations are needed to check which one is right.

In spite of the success of thermal models for hadrons produced in
the relativistic heavy-ion collisions, it is known that the
statistical model of multiparticle production leads to the
Boltzmann factor\cite{fermi,hagedorn,cyy} and can explain the
hadronic data produced in e$^+$e$^-$ and p$\bar{\mbox{p}}$
collisions. In the series of papers by T. T. Chou, C. N. Yang and
E. Yen~\cite{cyy} a statistical model for multiparticle production
is derived and applied successfully to the rapidity distribution
and transverse momentum distribution of measured hadrons in e$+$
e$^-$ collisions by TASSO collaboration. This model has been
further modified and studied by Hoang~\cite{hoang} to include the
Lorentz-boosted Boltzmann factor in order to mimic the shift of
the peak of the rapidity distribution in the asymmetric
collisions. It should be emphasized that this statistical model
does neither require the thermalization of the system nor the
concept of freeze-out. Recently the same argument has been
revived~\cite{hsu,rischke,koch}, but application of the
statistical model to the relativistic heavy-ion collision data has
not been attempted yet.

Thus it is very interesting to study whether the statistical model
which successfully fit e$^+$e$^-$ collision data can also fit the
rapidity distributions and transverse momentum spectra of pions
and kaons produced in the relativistic heavy-ion collisions,
especially whether it can fit all of them with a single set of
parameters. A priori it is not clear whether pions and kaons have
the same parameter "T" in the statistical model. However, there is
a claim that e$+$ e$^-$ collision data could be fit with only one
"temperature" for various hadrons\cite{cyy,hoang}.

In this paper we will show that a simple statistical model can
describe both the rapidity and transverse mass spectra of various
hadrons measured in the heavy-ion collisions. In our previous
work\cite{kslee}, the rapidity distribution of hadrons are
analyzed using the same model.

In the next section the statistical model by chou, Yang and Yen is
reviewed and in sec. III the result of our analysis of the
rapidity distributions and transverse mass spectra of $\pi^+$,
$\pi^-$, K$^+$ and K$^-$ in Pb+Pb collisions at 158 GeV$\cdot$ A
by the NA49 collaboration\cite{cooper,leeuwen} is presented and
finally we summarize in sec. IV.

%%%%%%%%%%%%%%%%%%%%%%%%%%%%%%%%%%%%%%%%%%%%%%%%%%%%%%%%
\section{Statistical Model }
\label{sec:model}
%%%%%%%%%%%%%%%%%%%%%%%%%%%%%%%%%%%%%%%%%%%%%%%%%%%%%%%%

In this section we follow the paper by Chou, Yang and
Yen~\cite{cyy} for the statistical model invented for e$^+$e$^-$
collisions which can be used for the heavy-ion collisions with
minor modifications.

We will assume the multiparticle production in the relativistic
heavy-ion collisions is stochastic but subject to a number of
conditions: (a) energy conservation, (b) leading baryon effect
which means that leading baryons in both the projectile and target
region do not participate in the collision, (c) $d^3 p/E$
probability for each particle, and (d) transverse momentum cutoff
factor $g(p_\perp )$. Then the probability for non-leading
particles in the collision zone can be taken as

\begin{equation}
 \delta~ (~\sum E_i - E_0 h~) \prod _i ~(d^3 p_i /E_i )~ g(p_{\perp _i} )
\end{equation}
where $E_0 = \sqrt{s}$ and $h$ is the fraction of the total energy
$E_0$ used for the particle production in the central region.

It was pointed out that the probability distribution is exactly
the same as in the microcanonical ensemble in the statistical
mechanics. By the well-known Darwin-Fowler method the single
particle distribution of such an ensemble can be converted into a
distribution in the canonical ensemble :

\begin{equation}
Probability = (d^3 p~ /E )~ g(p_\perp ) \exp (-E / T_p )
\end{equation}
where $T_p$ is called the partition temperature. Noting that $d^3
 p = dp_L p_{\perp} dp_{\perp} d\phi = E dy p_{\perp} dp_{\perp}
d\phi$, one gets the equation for the rapidity distribution by
integrating over the transverse momentum.

\begin{equation}
\label{eq:dndy}
 d N_i /d y = 2 \pi K \int _0 ^{p_{max}} p_{\perp} dp_{\perp}~
 g_i (p_\perp ) \exp (-E_i / T_p )
\end{equation}
where $K$ is a normalization constant, and $p_{max} = E_0 h$ which
can be taken as infinity in the relativistic heavy-ion collisions.
$E_i = m_T \cosh y$ , and $m_T = \sqrt{ m_i ^2 + p_{\perp} ^2}$.
We will use for the high momentum cutoff factor as in
Ref.~\cite{cyy},
\begin{equation}
\label{eq:gi}
 g_i(p_\perp ) = \exp (- \alpha _i p_{\perp})
\end{equation}
where $\alpha_i$ related to the average transverse momentum,
$<p_{\perp}>$. One can show that for the massless case,
 \begin{equation}
 \label{eq:alpha}
 \alpha_i = 2 [ <p_{\perp _i}> ]^{-1}
 \end{equation}
and in fitting e$^+$e$^-$ data, the value of $\alpha$ was taken
from the measured data.

Similarly the transverse momentum or transverse mass spectra is
given, after integration over the rapidity $y$, as
\begin{equation}
\label{eq:mtdmt}
 d N_i /m_{\perp} dm_{\perp} = 2 \pi K g_i(p_\perp ) \int d y  \exp (-E_i / T_p )
\end{equation}

Eq.~(\ref{eq:dndy}) can be integrated for the massless particles
and the role of parameters can be easily understood.
\begin{equation}
dN_i / dy =  \frac{2 \pi K }{(\alpha_i + \cosh y /T_p )^2} ,
\end{equation}
 whose maximum at $y=0$ is
\begin{equation}
(dN_i / dy)_{max} =  \frac{2 \pi K }{(\alpha_i + 1/T_p )^2} =
\frac{2 \pi K T_p ^2}{(\alpha_i  T_p +1)^2}
\end{equation}
and the width $\delta y_i$ at the half maximum is given from the
relation
\begin{equation}
\cosh \delta y_i = (\sqrt{2} -1) \alpha_i T_p + \sqrt{2}
\end{equation}
For the massless particle  the width is governed by the factor
$\alpha_i T_p $. Thus when the value $\alpha$ is taken from
experiment, the parameter $T_p$ is a measure of the width of
$dN/dy$ in this model. This is quite different from the thermal
models, where $T_p$ determines the slope of the transverse
momentum spectra and in order to fit the width of the rapidity
distribution one further need the concept of longitudinal
expansion of the system.

 This can be understood again from
Eq.~(\ref{eq:mtdmt}), which describes an exponential shape.
Contrary to the naive expectation, the slope of the exponential is
 determined not by $T_p$, but by $1/\alpha$ through Eq.~(\ref{eq:gi}).
This is because the fitted value of $T_p$ is very large ( $\sim $
1 GeV) compared to the average transverse momentum, which usually
translates into the temperature.  Thus in this model the main
slope of the exponential shape is put in by hand as a high
momentum cut-off factor, Eq.~(\ref{eq:gi}) and the integration in
Eq.~(\ref{eq:mtdmt}) gives a modification to the simple
exponential shape.

In Eq.~(\ref{eq:dndy}) and Eq.~(\ref{eq:mtdmt}) , $K$, $T_p$ and
$\alpha_i$ 's are parameters. In the next section results of
fitting of the rapidity distributions and transverse mass spectra
of $\pi^+$, $\pi^-$, K$^+$ and K$^-$ in Pb+Pb collisions at 158
GeV$\cdot$A by the NA49 collaborations.

%%%%%%%%%%%%%%%%%%%%%%%%%%%%%%%%%%%%%%%%%%%%%%%%%%%%%%%%
\section{Analysis }
\label{sec:fitting}
%%%%%%%%%%%%%%%%%%%%%%%%%%%%%%%%%%%%%%%%%%%%%%%%%%%%%%%%

Main results of the fitting of the rapidity distributions and
transverse mass spectra of $\pi^+$, $\pi^-$, K$^+$ and K$^-$ in
Pb+Pb collisions at 158 GeV$\cdot$A measured by NA49
collaboration\cite{cooper,leeuwen} are shown in Fig.~ 1-4 and the
resulting parameters are tabulated in Table 1.

In Table 1, we have fitted each species separately and thus the
resulting parameters are all different. $<p_{\perp} >$ is
calculated from $\alpha$ using Eq.~(\ref{eq:alpha}). One may use
experimental value of $\alpha$ and then $K$ and $T_p$ are the only
parameters in this model. It is amazing that  a very simple model
with only two parameters can fit both of the rapidity and
transverse momentum spectra of each particle species.

In this model there is no a priori reason for the common value of
$T_p$ for different particle species. Simultaneous fitting of the
rapidity and transverse mass spectra of pions and kaons with a
single value $T_p$ is not possible,  which is contrary to the case
of e$^+$ e$^-$ collisions\cite{hoang}. However, in Ref.
\cite{hoang} Lorentz-boosted exponential was used and direct
comparison is not possible and further studies with the
Lorentz-boosted exponential is needed to check whether in the
relativistic heavy-ion collisions a single value of $T_p$ can
describe the particle spectra of all the different species.

In Ref.~\cite{kslee} we have successfully fitted only the rapidity
distribution of pions and kaons with a single $T_p$ but different
$\alpha_i$'s. We get $K = 1212$, $T_p = 0.98 $ GeV,
$\alpha_{\pi^+} = 5.98$ c/GeV, $\alpha_{\pi^-} = 5.67$ c/GeV,
$\alpha_{K^+} = 11.4 $ c/GeV, and $\alpha_{K^-} = 17.1 $ c/GeV.
The large values of $\alpha_i$'s spoils the interpretation of
$\alpha$ in terms of the average transverse momentum through
Eq.~(\ref{eq:alpha}). This means that even though all the rapidity
distributions are fitted simultaneously, the transverse mass
spectra are far off. However, one can also fit the rapidity
distributions of each particle species separately by fixing the
$\alpha_i$ values as the experimental values and the resulting
values from the least square fit are approximately the same as
those in the Table 1, meaning that by fitting only the rapidity
distribution one can fit the transverse mass spectrum also.
Concentrating only for the rapidity distributions, one can either
fit them with a single value of $T_p$ with the transverse mass
spectra far off  or fit both of the rapidity and transverse mass
spectrum with different $T_p$ values for different particle
species.

As was discussed in the previous section $T_p$ is above 1 GeV and
thus in Eq.~(\ref{eq:mtdmt}), $1/\alpha$ plays the role of the
slope parameter and the integral in Eq.~(\ref{eq:mtdmt}) acts as
the high energy cut-off.

For pions we have used $K$, $T_p$, $\alpha_{\pi^+}$ and
$\alpha_{\pi^-}$ as parameters. One may use the same $T_p$ and
$\alpha$ but different $K$ values for $\pi^+$, $\pi^-$. Or one may
further introduce the charge chemical potential $\mu_c$ to write
$K_{\pi^+} = K \exp{( \mu_c/T_p )}$ and $K_{\pi^-} = K \exp{(
-\mu_c/T_p )}$. In this way one gets $K = 625.0$, $T_p = 1.21$
GeV, $\alpha = 4.17$ c/GeV and $\mu_c = 0.043 $ GeV. The quality
of the fitting is more or less the same. Similarly for kaons one
can introduce the strange chemical potential $\mu_s$ to write
$K_{K^+} = K \exp{( -\mu_s/T_p )}$ and $K_{K^-} = K \exp{(
\mu_s/T_p )}$. Then the rapidity and transverse spectra of both
the K$^+$ and K$^-$ can be fitted simultaneously with $K = 50.7$,
$T_p = 1.43$ GeV , $\alpha_{K^+} = 2.6$c/GeV,  $\alpha_{K^-} =
2.9$ c/GeV and $\mu_s = -0.415$ GeV.

\begin{table}
\caption{The fitting parameters of the rapidity and transverse
mass spectra}
\begin{tabular} {|c|c|r|c|c|r|}\hline
   & $K$ &  $T_p$  & $\alpha$    & $<p_{\perp}>$  &  $\chi^2 /n$
   \\
      &  & GeV & c/GeV &   GeV/c &
   \\
     \hline
$\pi^+$ & 599.9 & 1.27 & 4.23 & 0.47 &    5.4 \\ \hline

$\pi^-$ & 649.8 & 1.19 & 4.15 & 0.48 &   15.0 \\ \hline

$K^+$ & 66.86 & 1.57 & 2.69 & 0.74 &  2.5 \\ \hline

$K^-$ & 37.82 & 1.35 & 2.79 & 0.72 &   2.7 \\ \hline

\end{tabular}
\end{table}

%%%%%%%%%%%%%%%%%%%%%%%%%%%%%%%%%%%%%%%%%%%%%%%%%%%%%%%%
\section{Conclusion }

%%%%%%%%%%%%%%%%%%%%%%%%%%%%%%%%%%%%%%%%%%%%%%%%%%%%%%%%

We have fitted the rapidity and transverse mass spectra of
$\pi^+$, $\pi^-$, K$^+$ and K$^-$ in Pb+Pb collisions at 158
GeV$\cdot$A measured by NA49 collaboration with a simple
statistical model. For each particle species both the rapidity and
the transverse mass spectra are nicely fitted with only three
parameters. However, simultaneous fitting for all the pions and
kaons with a single value of $T_p$ is not possible.

Thus the success of thermal models for the hadron multiplicities
and/or transverse momentum spectra, by itself alone,  does not
necessarily mean the thermalization of the system.

\acknowledgments

We wish to acknowledge U. Heinz for useful
discussions and H. Dobler for his work in arranging data in useful
format.

%%%%%%%%%%% References %%%%%%%%%%%%%%%%%%%%%%%%%%%%%%%%%%%%%%%%%%%%%%%

%

%\newpage
\begin{figure}
\begin{center}
\epsfxsize=1.5in
 \epsffile{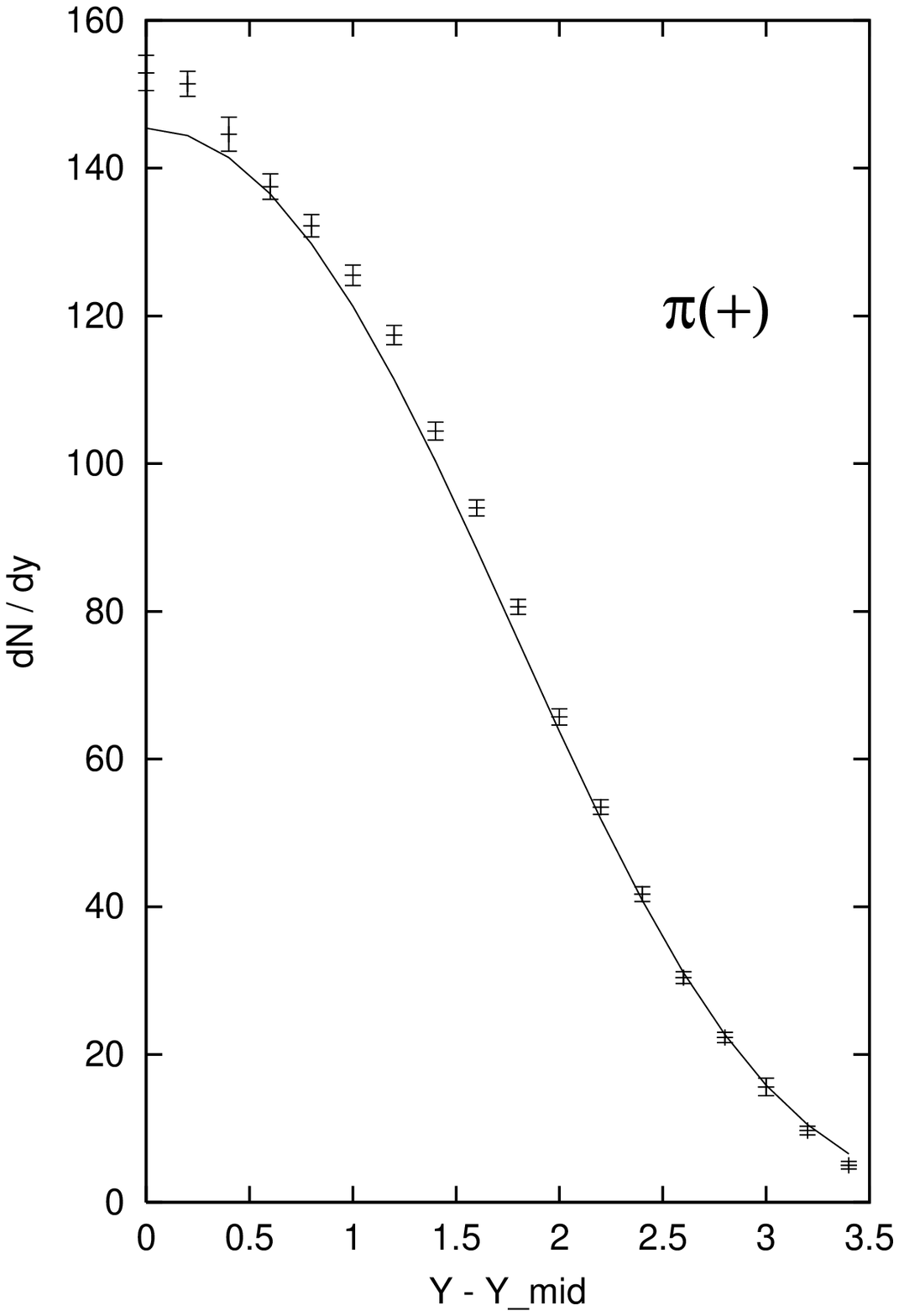} \hspace*{0.8cm}
\epsfxsize=1.5in
 \epsffile{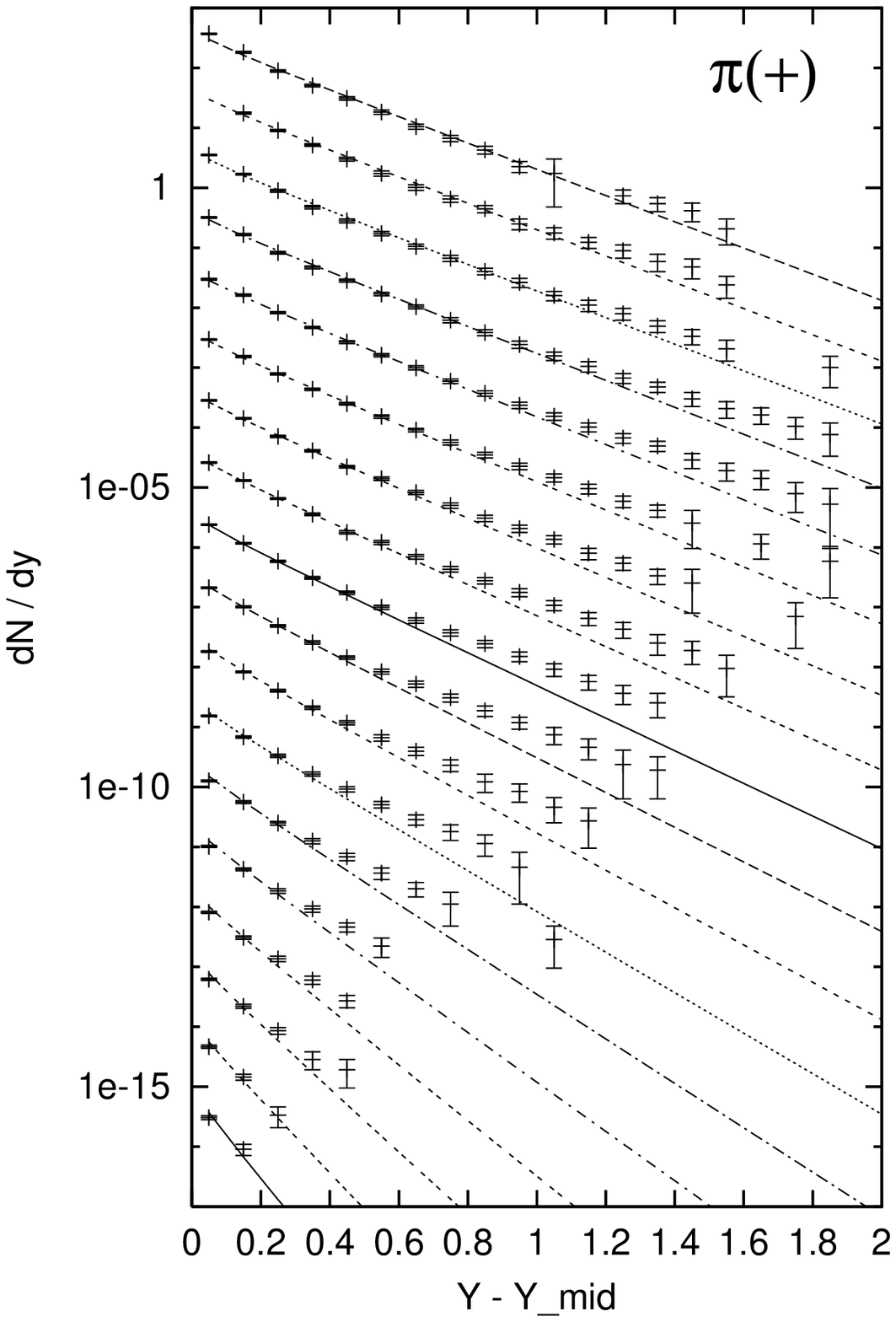}

 \caption{Rapidity distribution and transverse mass spectra of $\pi^+$
 in Pb+Pb collisions at 158 GeV$\cdot$A by NA49 collaborations. Lines are
  the fittings with Eq.~(\ref{eq:dndy}) and Eq.~(\ref{eq:mtdmt}) with parameters in
  Tab.~(1). In the right figure from the top  y bin is 0.0 $\pm
  0.1$ and the second one is for y =0.2$\pm$ 0.1, and so forth.
    }

\end{center}
  \label{fig1}
\end{figure}

\begin{figure}
\begin{center}
\epsfxsize=1.5in
 \epsffile{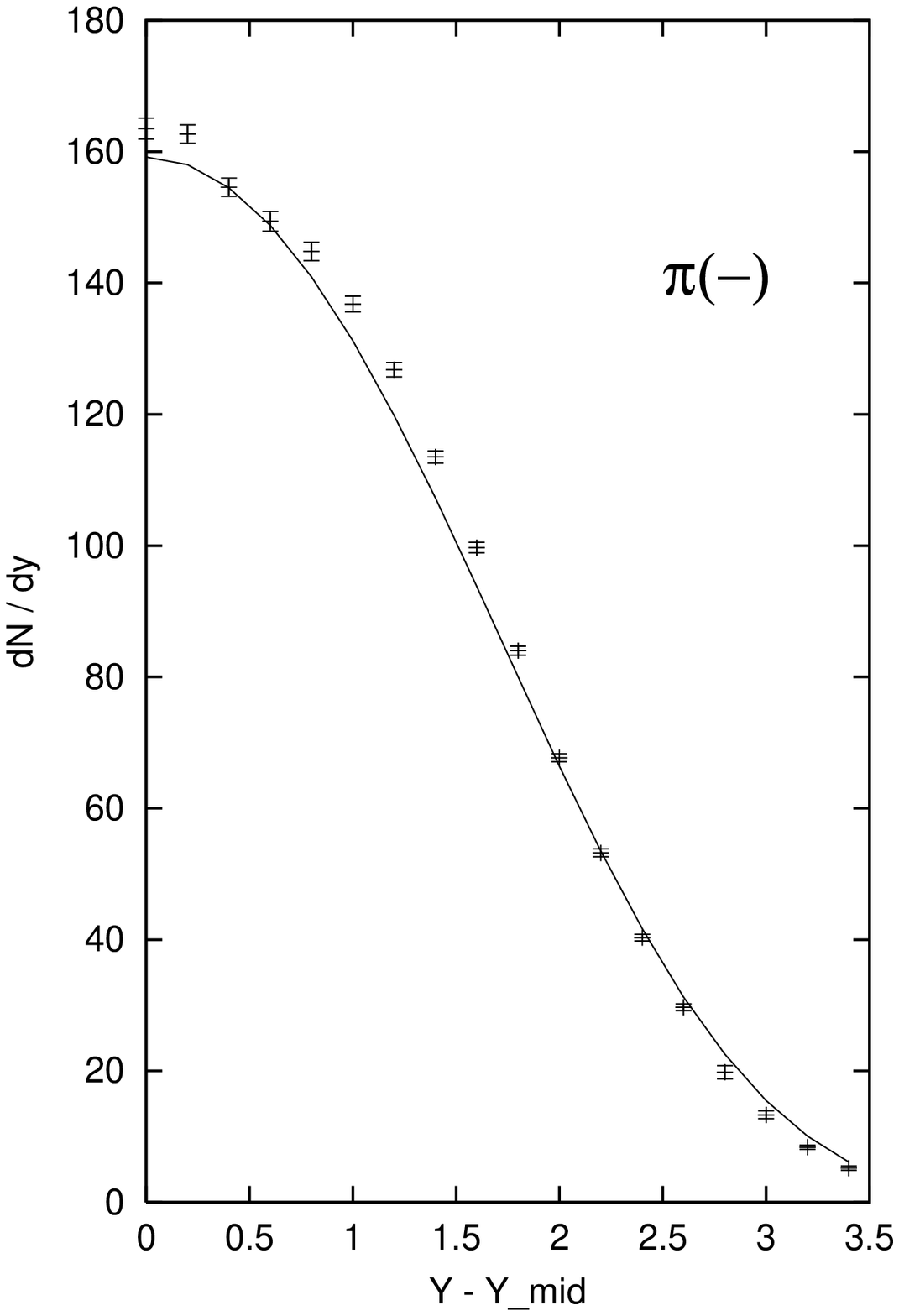} \hspace*{0.8cm}
\epsfxsize=1.5in
 \epsffile{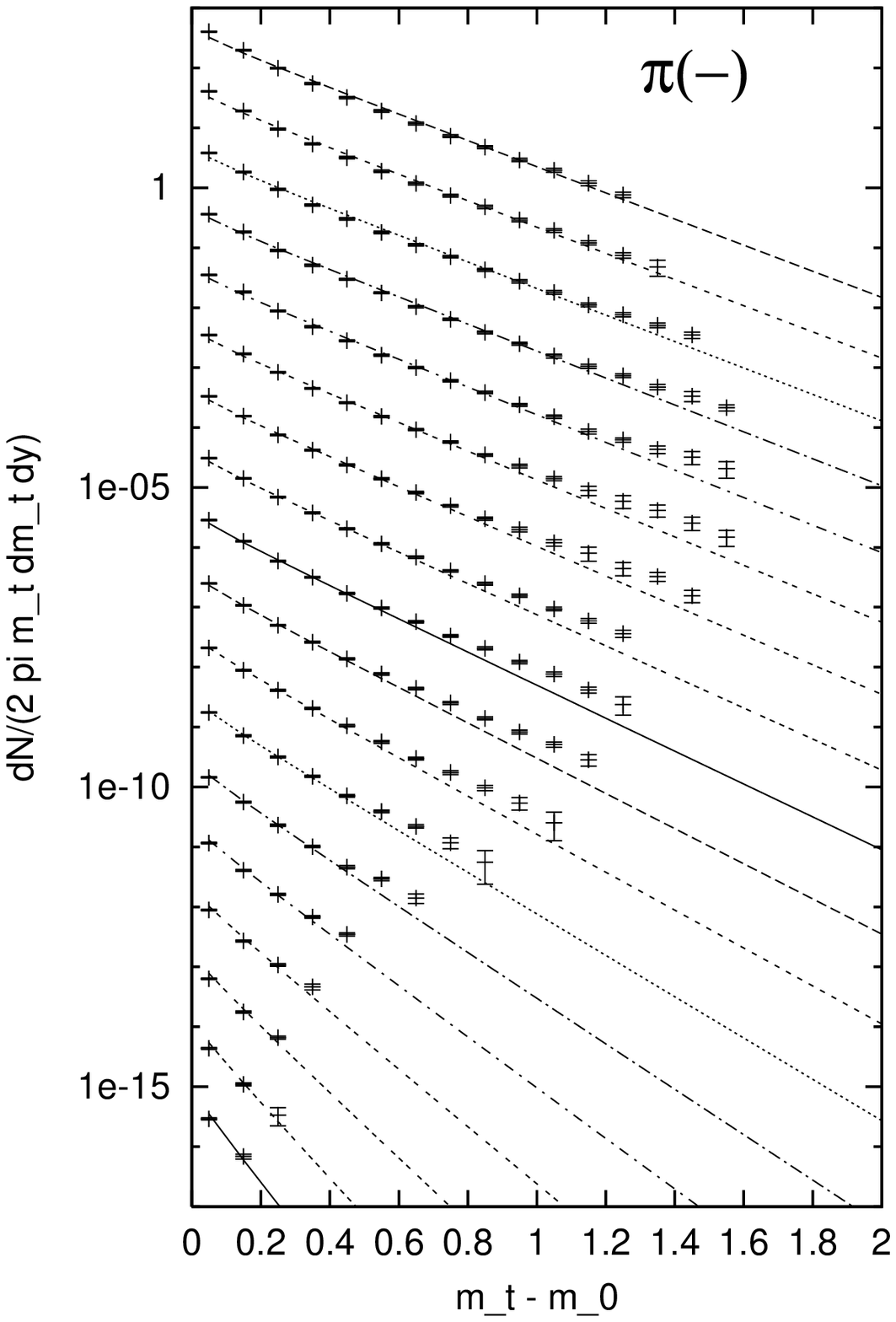}

 \caption{Same as in Fig1. for $\pi^- $.    }

\end{center}
  \label{fig2}
\end{figure}

\begin{figure}
\begin{center}
\epsfxsize=1.5in
 \epsffile{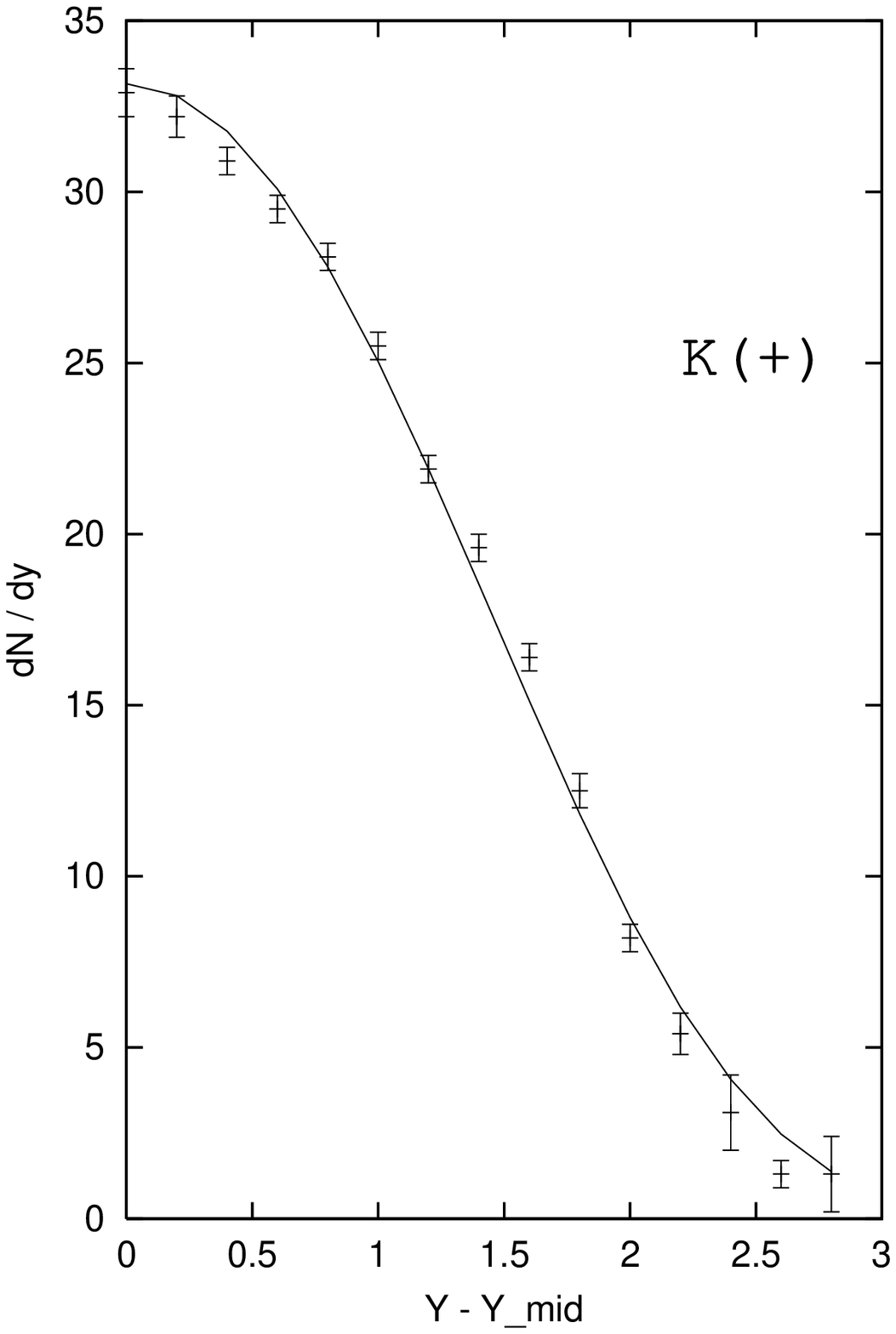} \hspace*{0.8cm}
\epsfxsize=1.5in
 \epsffile{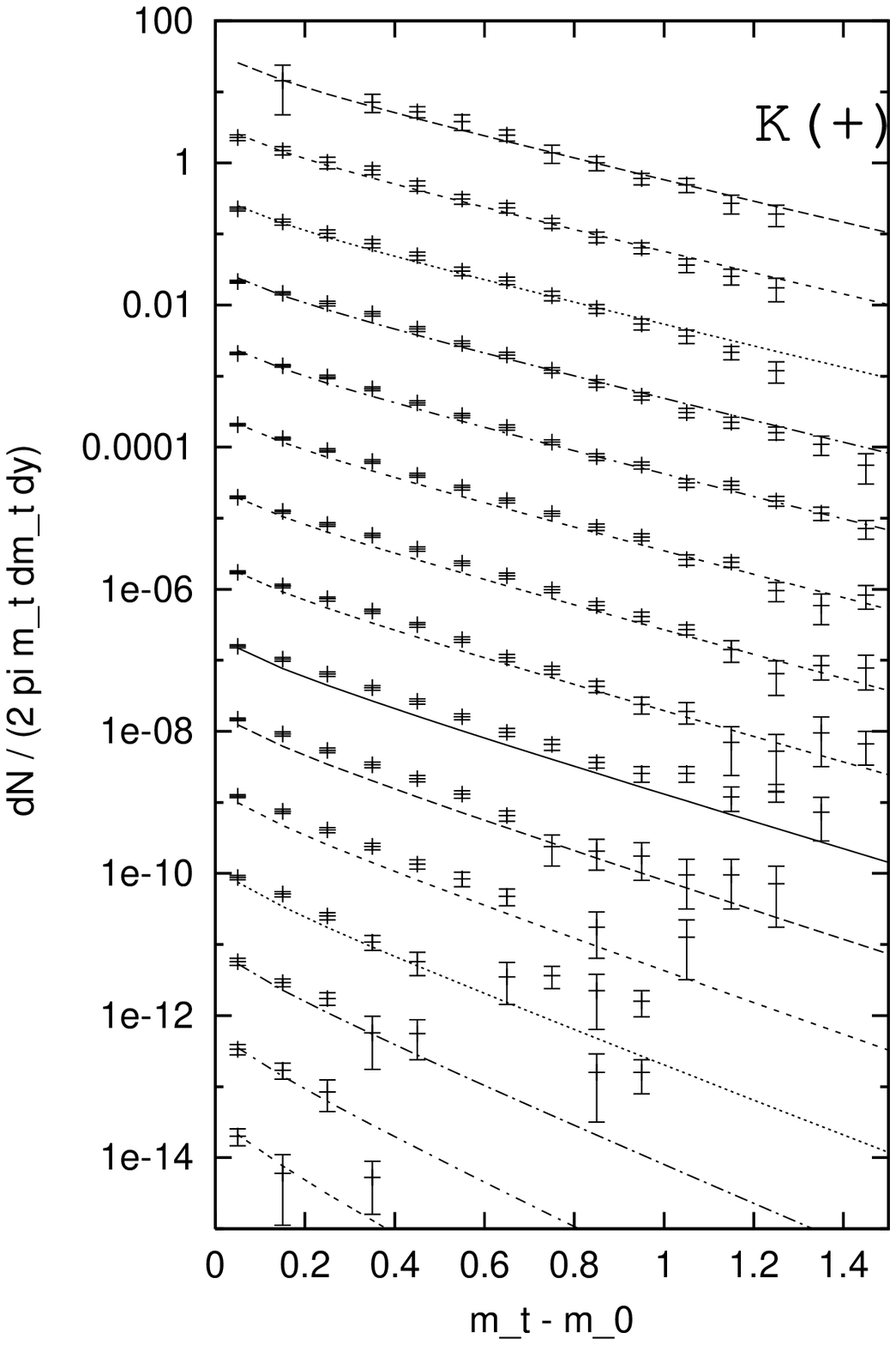}

 \caption{ Same as in Fig1. for K$^+$.     }

\end{center}
  \label{fig3}
\end{figure}

\begin{figure}
\begin{center}
\epsfxsize=1.5in
 \epsffile{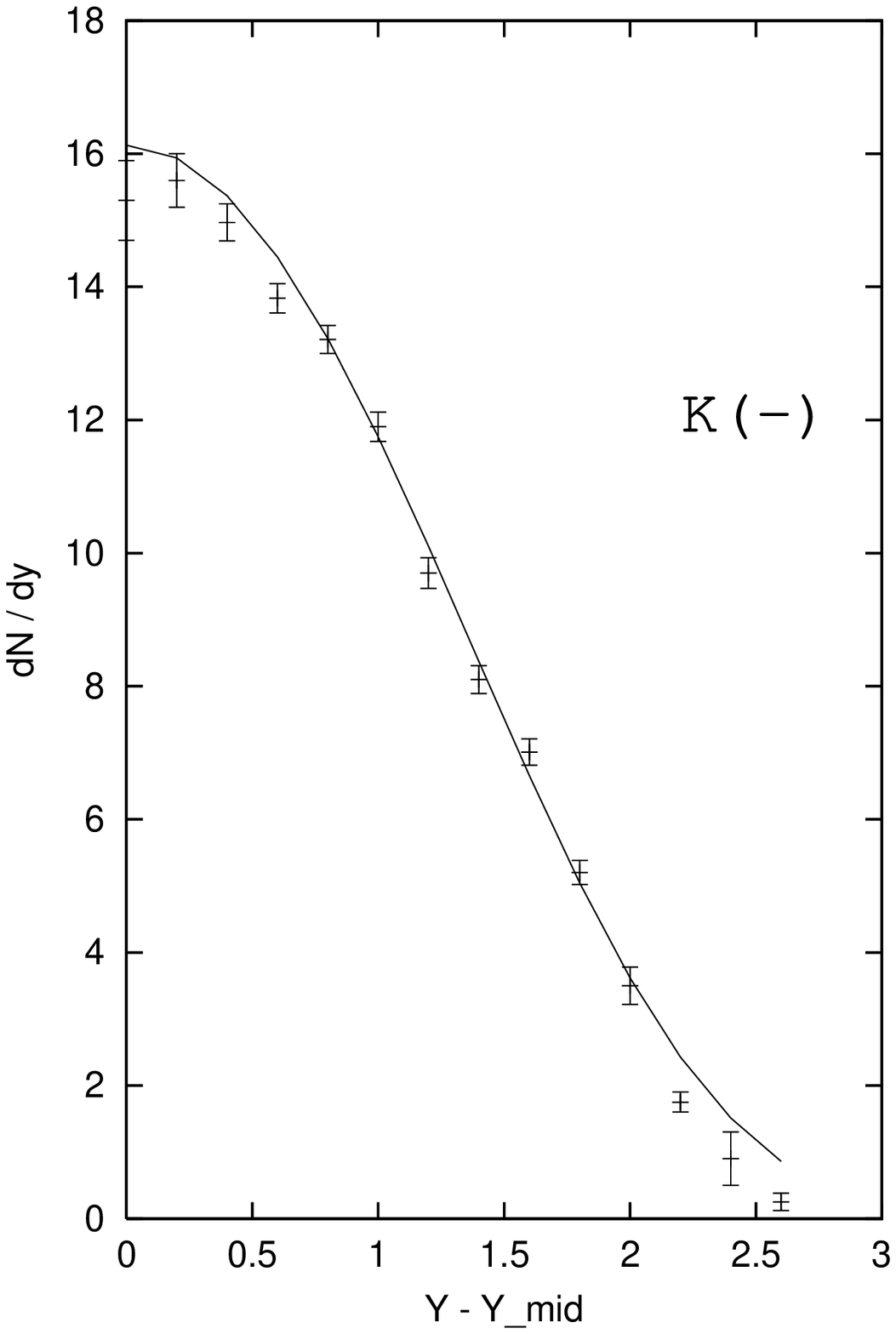} \hspace*{0.8cm}
\epsfxsize=1.5in
 \epsffile{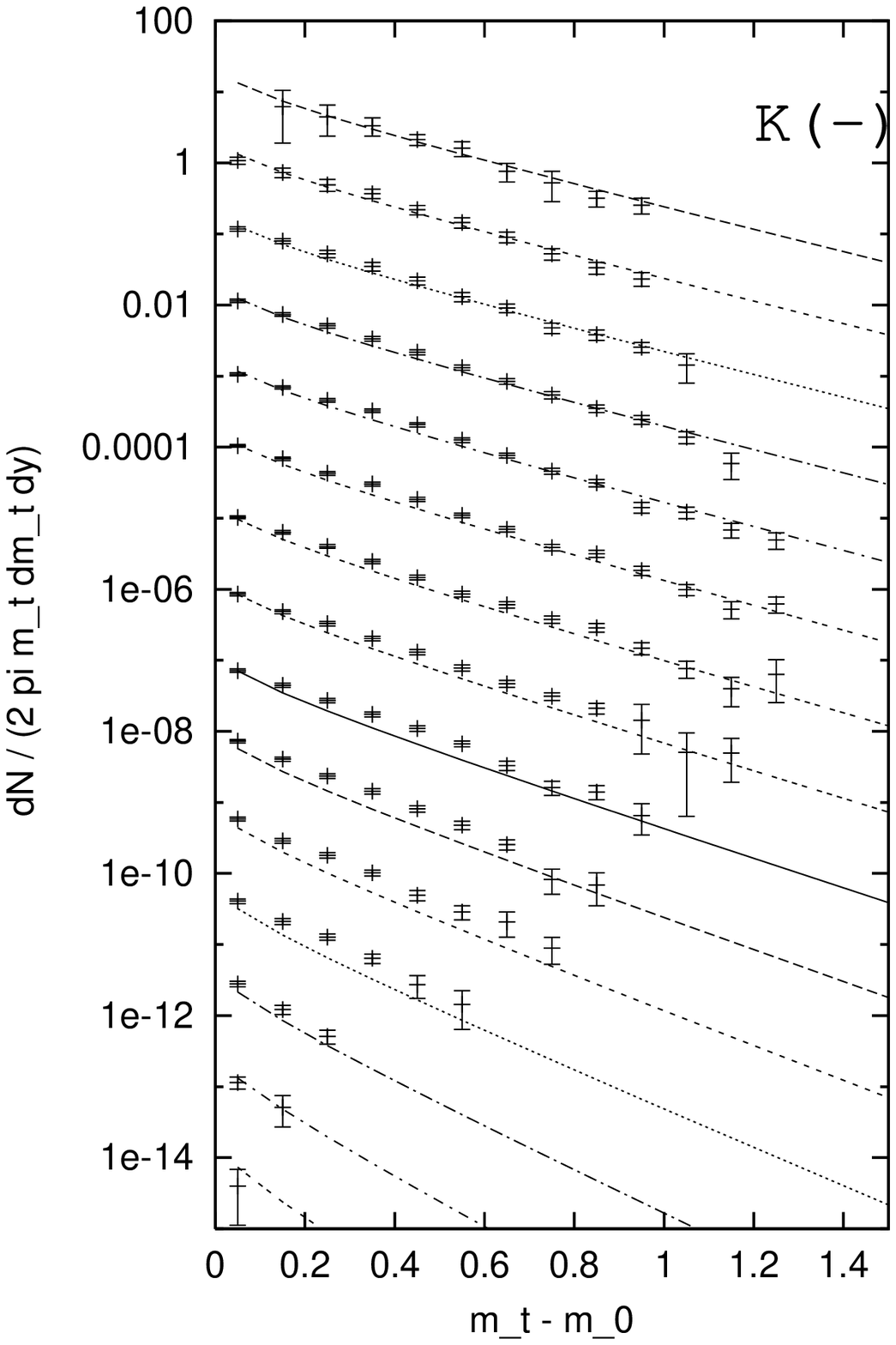}

 \caption{Same as in Fig1. for K$^-$.     }

\end{center}
  \label{fig4}
\end{figure}

%
%\end{multicols}

\end{document}